\documentclass[a4paper,11pt]{article}
\usepackage[skins]{tcolorbox}
\usepackage{mathtools,slashed}
\usepackage[T1]{fontenc}
\usepackage{slashed,verbatim,subfigure}
\usepackage[numbers,sort&compress]{natbib}
\usepackage{amsmath}
\usepackage{mathrsfs}
\usepackage{amsbsy}
\usepackage{amssymb}
\usepackage{appendix}
\usepackage{graphicx}
\usepackage[final]{pdfpages}
\usepackage{dcolumn}
\usepackage{bm}
\usepackage[colorlinks=true,linktocpage=true,
linkcolor=blue,citecolor=blue]{hyperref}
\usepackage[a4paper]{geometry}
\catcode`@11
\def\seceqaa{\@addtoreset{equation}{section}
	\def\theequation{A\arabic{equation}}}
\def\seceqbb{\@addtoreset{equation}{section}
	\def\theequation{B\arabic{equation}}}
\def\seceqcc{\@addtoreset{equation}{section}
	\def\theequation{C\arabic{equation}}}
\def\seceqdd{\@addtoreset{equation}{section}
	\def\theequation{D\arabic{equation}}}
\def\seceqee{\@addtoreset{equation}{section}
	\def\theequation{E\arabic{equation}}}
\catcode`@11
\newcommand{\be}{\begin{eqnarray}}
\newcommand{\ee}{\end{eqnarray}}
\begin{document}
\large
\title{Page Curves of Reissner-Nordstr{\"o}m Black Hole in HD Gravity}
\author{Gopal Yadav\footnote{email- gyadav@ph.iitr.ac.in} \vspace{0.1in}\\
Department of Physics,\\
Indian Institute of Technology Roorkee, Roorkee 247667, India}
\date{}
\maketitle
\begin{abstract}
We obtain the Page curves of an eternal Reissner-Nordstr{\"o}m black hole in the presence of higher derivative terms in four dimensions. We consider two cases: gravitational action with general ${\cal O}(R^2)$ terms plus Maxwell term and Einstein-Gauss-Bonnet gravity plus Maxwell term. In both the cases entanglement entropy of the Hawking radiation in the absence of island surface is increasing linearly with time. After including contribution from the island surface, we find that after the Page time, entanglement entropy of the Hawking radiation in both the cases reaches a constant value which is the twice of the Bekenstein-Hawking entropy of the black hole and we obtain the Page curves. We find that Page curves appear at later or earlier time when the Gauss-Bonnet coupling increases or decreases. Further, scrambling time of  Reissner-Nordstr{\"o}m is increasing or decreasing depending upon whether the correction term (coming from ${\cal O}(R^2)$ terms in the gravitational action)  is increasing or decreasing in the first case whereas scrambling time remains unaffected in the second case (Einstein-Gauss-Bonnet gravity case). As a consistency check, in the limit of vanishing GB coupling we obtain the Page curve of the Reissner-Nordstr{\"o}m black hole obtained in \cite{Island-RNBH}.
\end{abstract}

\newpage
\tableofcontents
\newpage
\section{Introduction and Motivation}
\label{I-M}
The black hole information paradox \cite{Hawking} has to do with the unitary evolution of the black holes. Page proposed that entropy of Hawking radiation of an evaporating black hole should fall after the Page time \cite{Page}. For eternal black holes entropy of the Hawking radiation should reach a constant value which is Bekenstein-Hawking entropy \cite{BH-entropy} of the black holes. To calculate the Page curve of a black hole one is required to couple the black hole to bath where we can collect the Hawking radiation. For eternal black holes we take two copies of the aforementioned setup to calculate the Page curve. 
\par 
With the discovery of AdS/CFT correspondence \cite{AdS-CFT} as a duality between conformal field theory and anti-de Sitter background in one higher dimensions, it become easy to resolve the black hole information paradox \cite{Hawking}. We can obtain the entanglement entropy of conformal field theory from gravity dual background using Ryu-Takayanagi formula \cite{RT} for static background. For time dependent background one is required to use the prescription given in \cite{HRT}. Quantum corrections to all order in $\hbar$ to the Ryu-Takayanagi formula was incorporated in \cite{EW} where one is required to extremize the generalised entropy. Surfaces which extremize the generalised entropy are known as quantum extremal surfaces(QES). If there are more than one quantum extremal surfaces then we need to consider the one with minimal area. In \cite{AMMZ}, authors generalised the QES prescription to island surfaces where we are required to extremize the generalised entropy like functional which includes contribution from the island surfaces. In this case extremal surfaces are known as quantum extremal islands.
\par
We can calculate the Page curves of eternal black holes using Island proposal \cite{AMMZ}. Island formula is given below:
\begin{eqnarray}
\label{Island-proposal}
S_{\rm gen}(r)={\rm Min}_{\cal I} \Biggl[ {\rm Ext}_{\cal I}\Biggl(\frac{Area(\partial {\cal I})}{4 G_N} + S_{\rm matter}({\cal R} \cup {\cal I})\Biggr)\Biggr],
\end{eqnarray}
where ${\cal R}$ is the radiation region, $G_N$ is the Newton constant and ${\cal I}$ is the island surface. First term in the above formula is the area of the island surface and second term is the matter contribution to the entanglement entropy of the Hawking radiation which includes contribution from the island surface. Since initially there is no island surface therefore (\ref{Island-proposal}) reduces to the entanglement entropy of only radiation part which increases linearly with time. At late times island surface comes into the picture and saturates the entanglement entropy growth of the Hawking radiation at the Page time and produces the Page curve. In general higher derivative gravity we can write the above formula in most general form as,
\begin{eqnarray}
\label{S-total-general}
S_{\rm total} = S_{\rm gravity}+ S_{\rm matter},
\end{eqnarray}
where $S_{\rm gravity}$ and $S_{\rm matter}$ are the gravitational and matter contributions to the total entanglement entropy. In higher derivative gravity theories first term in equation (\ref{S-total-general}) can be calculated using Dong formula \cite{Dong} and  $S_{\rm matter}$ can be calculated using Cardy formula \cite{CC,CC-1}. Then we are required to extremize the total entanglement entropy with respect to location of the island surfaces. If there are more than one surface then we need to choose the surface with the minimal area among those surfaces. By following this approach Page curves of the Reissner-Nordstr{\"o}m black hole, charged dilaton black hole, Schwarzschild black hole and hyperscaling violating black branes were calculated in \cite{Island-RNBH,Yu-Ge,Island-SB,Omidi}. Page curve in charged linear dilaton model for non-extremal black hole and the extremal black hole have been studied in \cite{CLDBH}\footnote{We thank H.~S.~Jeong to bring their work to our attention.}. Recently islands in Kerr-de Sitter spacetime and generalized dilaton theories have been studied in \cite{Islands-KdS,Tian}. Page curve and the information paradox for the flat space black holes was studied by authors in \cite{Flat-space-black-holes}\footnote{We thank C.~Krishnan to bring their work to our attention.}. The role played by mutual information of subsystems on the Page curve was studied in \cite{M-Page}\footnote{We thank A. Saha to bring their work to our attention.}.
Page curve calculations in higher derivative gravity theories can be found in \cite{NBH-HD,HD-Page-curve-2}. In \cite{NBH-HD}, authors calculated Page curves of the Schwarzschild black holes in the presence of higher derivative terms which are ${\cal O}(R^2)$ terms. By following \cite{NBH-HD}, we are calculating the Page curves of an eternal Reissner-Nordstr{\"o}m black hole in the presence of  ${\cal O}(R^2)$ terms as considered in \cite{NBH-HD} and in Einstein-Gauss-Bonnet gravity \cite{Charged-GB-BH} in four dimensions in this paper.
 \par
Charged black hole in Einstein-Gauss-Bonnet gravity in four dimensions were studied in \cite{Charged-GB-BH-App,Zhang-Li-Guo,BBB,ZZZY,CGLY,BB,LNZ}. In \cite{Zhang-Li-Guo}, authors studied superradiance, stability and quasinormal modes of regularised charged Einstein-Gauss-Bonnet black hole. In \cite{BBB}, authors have discussed thermodynamics of the charged Einstein-Gauss-Bonnet black hole via van der Walls equation. In \cite{ZZZY}, thermodynamics and phase transitions of charged AdS black holes was studied, additionally quasinormal modes for the massless scalar perturbation was also studied in the same paper. In \cite{CGLY}, correspondence between shadow and test field was studied for the charged Einstein-Gauss-Bonnet black hole. In \cite{BB}, charge of the Einstein-Gauss-Bonnet black hole was studied by considering negative and positive charge of the particle-antiparticle pair on cauchy horizon and physical horizon. In \cite{LNZ}, instability of charged Einstein-Gauss-Bonnet de-Sitter black holes was studied by authors under charged scalar perturbations. In \cite{Charged-GB-BH-App}, authors have discussed motion of the charged and spinning particles and photons in the vicinity of charged Einstein-Gauss-Bonnet black hole in four dimensions. Further radius of the innermost circular orbit, gravitational deflection angle and some collision processes were also studied by the authors.  For review on Einstein-Gauss-Bonnet gravity in four dimensions, see \cite{EGB-Review}.
\par 
As summarised above literatures on the charged Einstein-Gauss-Bonnet black hole in four dimensions. We find that study of effect of the Gauss-Bonnet coupling on the Page curve of charged Einstein-Gauss-Bonnet black hole was missing. Therefore
it will be interesting to study that how the Page curves of Reissner-Nordstr{\"o}m black hole will be modified with the presence of Gauss-Bonnet term. With this motivation we have calculated the Page curves of charged black hole in higher derivative gravity with ${\cal O}(R^2)$ terms and for the charged Einstein-Gauss-Bonnet black hole in this paper.
\par
Structure of the paper is as follow. Section {\ref{review}} has been divided into two subsections {\ref{review-CEGB-BH}} and {\ref{review-RNBH-Page-curve}}. In subsection {\ref{review-CEGB-BH}} we review the charged black hole solution in Einstein-Gauss-Bonnet gravity in four dimensions \cite{Charged-GB-BH} and in subsection {\ref{review-RNBH-Page-curve}} we review the Page curve calculation of the Reissner-Nordstr{\"o}m black hole \cite{Island-RNBH}. In section {\ref{Page-curve-R^2}} we have calculated the Page curves of an eternal Reissner-Nordstr{\"o}m black hole in the presence of ${\cal O}(R^2)$ terms in the gravitational action. In section {\ref{Page-curves-CEGB-BH}} we have calculated the Page curves of the same black hole in Einstein-Gauss-Bonnet gravity. In section {\ref{Summary}} we have some discussion about the island rule which also includes the summary of the results obtained in this paper.

\section{Review}
\label{review}
We have divided this section into two subsections.
In subsections {\ref{review-CEGB-BH} and {\ref{review-RNBH-Page-curve}, we are briefly reviewing the charged black hole in Einstein-Guass-Bonnet gravity in the vanishing cosmological constant limit \cite{Charged-GB-BH} and Page curve calculation of Reissner-Nordstr{\"o}m black hole \cite{Island-RNBH} in four dimensions.
\subsection{Brief Review of Charged Black Hole in Einstein-Gauss-Bonnet Gravity}
\label{review-CEGB-BH}
In this section we are reviewing spherically symmetric charged black hole solution in Einstein-Gauss-Bonnet gravity in four dimensions \cite{Charged-GB-BH}. Authors in \cite{Glavan+Lin} have shown that we can make the Gauss-Bonnet term in four dimensions dynamical if we rescale the Gauss-Bonnet coupling as $\alpha \rightarrow \frac{\alpha}{D-4}$. Consistent theory of Gauss-Bonnet gravity was constructed by authors in \cite{AGM}\footnote{We thank S.~Siwach to bring \cite{AGM} to our attention.} in $(d+1)$-dimensions where all the relevant degrees of freedom is present.
Action of the  Aoki, Gorji and Mukhohyama (AGM) theory is:
\begin{eqnarray}
\label{action-AGM}
S=\frac{1}{2 \kappa^2}\int d^{(d+1)}x \sqrt{-g}\left(R[g]+\alpha {\cal L}_{\rm GB}\right),
\end{eqnarray} 
where ${\cal L}_{\rm GB}=R_{\mu\nu\rho\sigma}[g]R^{\mu\nu\rho\sigma}[g]-4 R_{\mu\nu}[g]R^{\mu\nu}[g]+R[g]^2$ is the Gauss-Bonnet term. If we take the $d \rightarrow 3$ limit of equation (\ref{action-AGM}) then we obtain the consistent theory of Gauss-Bonnet gravity in four dimensions for the neutral black hole.
Author in \cite{Charged-GB-BH} constructed charged Einstein-Gauss-Bonnet black hole solution in four dimensions by rescaling of the Gauss-Bonnet coupling. We are discussing this solution in the vanishing cosmological constant limit. Action of the Einstein-Gauss-Bonnet gravity including Maxwell term is:
\begin{eqnarray}
\label{action-GB}
S=\frac{1}{2 \kappa^2}\int d^4x \sqrt{-g}[R[g]+\alpha {\cal L}_{\rm GB}-F_{\mu \nu}F^{\mu \nu}],
\end{eqnarray}
where, $R[g]$ is the Ricci scalar, $\alpha$ is the Gauss-Bonnet coupling, ${\cal L}_{\rm GB}$ is the Gauss-Bonnet term and $F_{\mu\nu}$ is the field strength tensor corresponding to $A_\mu$ gauge field which are defined below:
\begin{eqnarray}
\label{LGB-F}
& &
{\cal L}_{\rm GB}=R_{\mu\nu\rho\sigma}[g]R^{\mu\nu\rho\sigma}[g]-4 R_{\mu\nu}[g]R^{\mu\nu}[g]+R[g]^2, \nonumber\\
& & F_{\mu\nu}=\partial_\mu A_\nu -\partial_\nu A_\mu.
\end{eqnarray}
Black hole solution of the action (\ref{action-GB}) is:
\begin{eqnarray}
\label{metric-GB}
ds^2=-F(r)dt^2+\frac{dr^2}{F(r)}+r^2 d\Omega^2,
\end{eqnarray}
where,
\begin{equation}
\label{F(r)}
F(r)= 1+\frac{r^2}{2 \alpha}\left(1 \pm\sqrt{1+4\alpha\left(\frac{2 M}{r^3}-\frac{Q^2}{r^4}\right)}\right).
\end{equation}
If we perform small expansion in $\alpha$ with negative sign chosen in (\ref{F(r)}) then we obtain:
\begin{equation}
\label{Fr-small-alpha}
F(r)=1-\frac{2 M}{r}+\frac{Q^2}{r^2}+\frac{\left(Q^2-2Mr\right)^2}{r^6}\alpha.
\end{equation}
Black hole horizons are given by solving $F(r)=0$ for the negative sign in equation (\ref{F(r)}), which are given below:
\begin{eqnarray}
\label{horizons}
r_{\pm}=M\pm \sqrt{M^2-Q^2-\alpha},
\end{eqnarray}
where $r_+$ is the physical horizon of the charged black hole and $r_-$ is the Cauchy horizon. Now let us see how the black hole horizon $r_+$ is varying with the Gauss-Bonnet coupling $\alpha$. We have plotted it in figure {\ref{rplus-alpha-variation}} for $M=1$ and various values of the black hole charges.

\begin{figure}
\begin{center}
\includegraphics[width=0.60\textwidth]{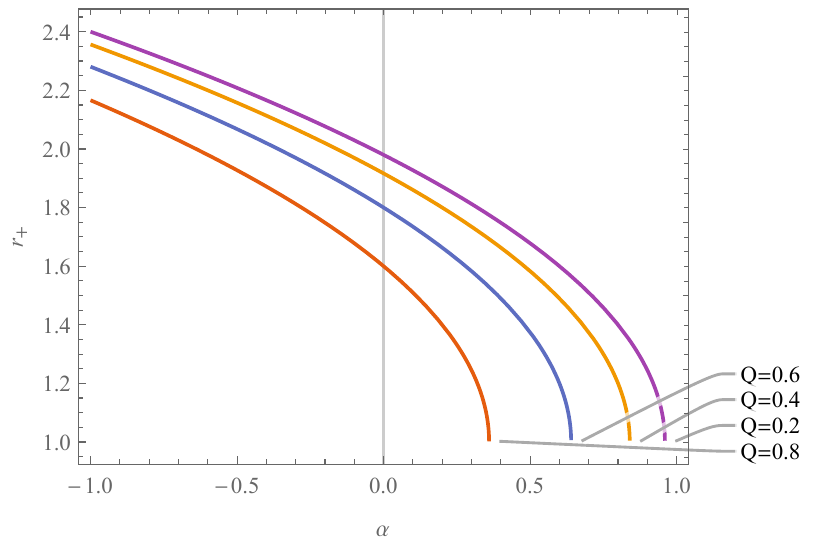}
\end{center}
\caption{Plot of $r_+$ versus $\alpha$}
\label{rplus-alpha-variation}
\end{figure}
From the figure {\ref{rplus-alpha-variation}} and equation (\ref{horizons}), we can see that when Gauss-Bonnet coupling ($\alpha$) is zero then we have usual Reissner-Nordstr{\"o}m black hole. When  Gauss-Bonnet coupling is decreasing towards negative values then black horizon is increasing and when Gauss-Bonnet coupling is increasing towards positive values then black horizon is decreasing.\par 
Authors in \cite{Zhang-Li-Guo,CGLY} have discussed that for $M=1$ following are the allowed values of the Gauss-Bonnet coupling $\alpha$.
\begin{itemize}
\item {\bf A:}$Q^2-4-2\sqrt{4-2Q^2}< \alpha < 1-Q^2, \ \ {\rm when} \ \ 0<Q<\frac{\sqrt{3}}{2}$.

\item {\bf B:}$Q^2-4-2\sqrt{4-2Q^2}< \alpha <Q^2-4+2\sqrt{4-2Q^2}, \ \ {\rm when} \ \ 0<Q<\sqrt{2}$.
\end{itemize}

As discussed in \cite{Zhang-Li-Guo,CGLY}, {\bf A} region contain both the horizons ($r_{\pm}$) whereas  {\bf B} region contain only the physical horizon ($r_+$). Since we are interested in the non-extremal black holes therefore relevant region for us is the region {\bf A}. Numerically in region {\bf A}, $\alpha \in (-6.66,0.36)$ for $Q=0.8$. Authors in \cite{SGB-1,SGB-2} studied the effect of Gauss-Bonnet term on the shear viscosity to entropy density ratio $\left(\frac{\eta}{s}\right)$, causality violation and instability of the black brane in five dimensions and arbitrary $D$ dimensions from the perspective of gauge-gravity duality. They found that there is an upper bound on the Gauss-Bonnet coupling, $\lambda \le \frac{1}{4}(\lambda \propto \alpha^{'}$ via $\lambda =\frac{(D-4)(D-3) \alpha^{'}}{l^2}$: $\alpha^{'}$ appear as Gauss-Bonnet coupling in \cite{SGB-1,SGB-2}) when $D \rightarrow \infty$ and, $\lambda \le 0.09$ when $D \rightarrow 5$. There is instability of RN-AdS black brane in five dimensions when $0<\lambda \le 0.09$.  
\par
Charged black hole in Einstein-Gauss-Bonnet gravity were studied in detail in various papers mentioned earlier in the introduction section {\ref{I-M}}. In this paper we will calculate the Page curves of the charged black hole in the presence of higher derivative terms which are ${\cal O}(R^2)$ terms and then we focus explicitly on the Einstein-Gauss-Bonnet gravity \cite{Charged-GB-BH}. Since we are working with higher derivative gravity therefore we will use \cite{Dong} to calculate the entanglement entropy wherever required for higher derivative terms in this paper.
\par
In this paper we are focusing only on the non-extremal black holes. Page curve of the extremal black holes were studied in \cite{Kim+Nam,CLDBH,Yu et al}. Authors in \cite{CLDBH,Kim+Nam} studied the Page curve of eternal charged linear dilaton black holes and Reissner-Nordstrom black holes for the non-extremal and extremal both cases and found that island formulation is ill-defined for the extremal black holes, which implies that we may not obtain the Page curve for the extremal black holes. Authors in \cite{Yu et al} considered rotating BTZ(Ba$\tilde{n}$ados-Teitelboim-Zanelli) black holes in three dimensions and found that for the non-extremal rotating BTZ black holes one obtains the Page curve using island formulation where as for the extremal rotating BTZ black holes, Page time and scrambling time turn out to be divergent and this problem can be removed by considering superradiance phenomenon in the system. With the inclusion of superradiance, Page time and scrambling time were decreasing as the angular momentum of the black holes were increasing and interestingly one obtains the Page curve after finishing of superradiance.

\subsection{Review of Page Curve of Reissner-Nordstr{\"o}m Black Hole}
\label{review-RNBH-Page-curve}
In this section we are reviewing the Page curve calculation of Reisnner-Nordstr{\"o}m black hole done in \cite{Island-RNBH}.
Action for the Einstein-Maxwell system is:
\begin{eqnarray}
I=\frac{1}{16 \pi G_N} \int_{\cal M} d^4x\sqrt{-g}\left(R-\frac{1}{4}F_{\mu\nu}F^{\mu\nu}\right)+I_{\rm matter},
\end{eqnarray}
where $R$ is the Ricci scalar, $F_{\mu\nu}$ is the field strength tensor corresponding to the $A_\mu$ gauge field and $I_{\rm matter}$ is the matter contribution to the action. Metric of the Reissner-Nordst{\"o}rm black hole is:
\begin{eqnarray}
\label{metric-RNBH-original}
ds^2=-F(r)dt^2+\frac{dr^2}{F(r)}+r^2 d\Omega^2,
\end{eqnarray}
where $d\Omega^2=d\theta^2+ \sin^2\theta d\phi^2$ and,
\begin{equation}
\label{F(r)-RNBH-original}
F(r)=1-\frac{2 M}{r}+\frac{Q^2}{r^2}.
\end{equation}
Black hole horizons will be given by solution to the $F(r)=0$, and are given below:
\begin{eqnarray}
\label{horizons-RNBH-original}
r_{\pm}=M\pm \sqrt{M^2-Q^2}.
\end{eqnarray}
We can write metric (\ref{metric-RNBH-original}) in Kruskal coordinates as given below \cite{PM}:
\begin{eqnarray}
\label{metric-kruskal-coordinates}
ds^2=-\frac{r_+ r_-}{r^2 \kappa_+^2}\left(\frac{r_-}{r-r_-}\right)^{\frac{\kappa_+}{\kappa_-}-1} e^{-2 \kappa_+ r} dUdV+r^2 d\Omega^2,
\end{eqnarray}
where 
\begin{equation}
\label{U-V}
U = - e^{-\kappa_+(t-r_*)}, V=e^{\kappa_+(t+r_*)},
\end{equation}
 and $r_*$ is defined as:
\begin{eqnarray}
\label{rstar-U}
& & r_*=r+\frac{r_+^2}{r_+ - r_-} \log|r-r_+|-\frac{r_-^2}{r_+ - r_-} \log|r-r_-|.
\end{eqnarray}
$\kappa_\pm$ are the surface gravities corresponding to $r_+$ and $r_-$ and are defined as:
\begin{eqnarray}
\label{surface-gravity}
& & \kappa_{\pm} =\frac{r_{\pm}-r_{\mp}}{2 r_{\pm}^2}.
\end{eqnarray}
Hawking temperature and Bekenstein-Hawking entropy of the charged black hole are defined below:
\begin{eqnarray}
& & T_{\rm RN}=\frac{\kappa_+}{2 \pi}, \nonumber\\
& & S_{\rm BH}^{(0)} = \frac{\pi r_+^2}{G_N}.
\end{eqnarray}
Defining the conformal factor in the metric (\ref{metric-kruskal-coordinates}) as:
\begin{eqnarray}
\label{conformal-factor}
g^2(r)=\frac{r_+ r_-}{r^2 \kappa_+^2}\left(\frac{r_-}{r-r_-}\right)^{\frac{\kappa_+}{\kappa_-}-1} e^{-2 \kappa_+ r}.
\end{eqnarray}
Therefore we can write the metric (\ref{metric-kruskal-coordinates}) in the following form:
\begin{eqnarray}
ds^2=-g^{2}(r) dUdV+r^2 d\Omega^2.
\end{eqnarray}
Penrose diagrams of an eternal Reissner-Nordstr{\"o}m black hole are given in figures \ref{p1} and \ref{p2} where $R_{-}$ and $R_+$ are the left and right wedges of the radiation regions, $b_{-}$ and $b_+$ are the boundaries of the $R_{-}$ and $R_+$, and, $a_{-}$ and $a_+$ are the boundaries of the island surface in the left and right wedges.\footnote{The author is thankful to X.~Wang for allowing to use the figures from their paper.}.

\begin{figure}
\centering
\begin{minipage}{.5\textwidth}
  \centering
  \includegraphics[width=.7\linewidth]{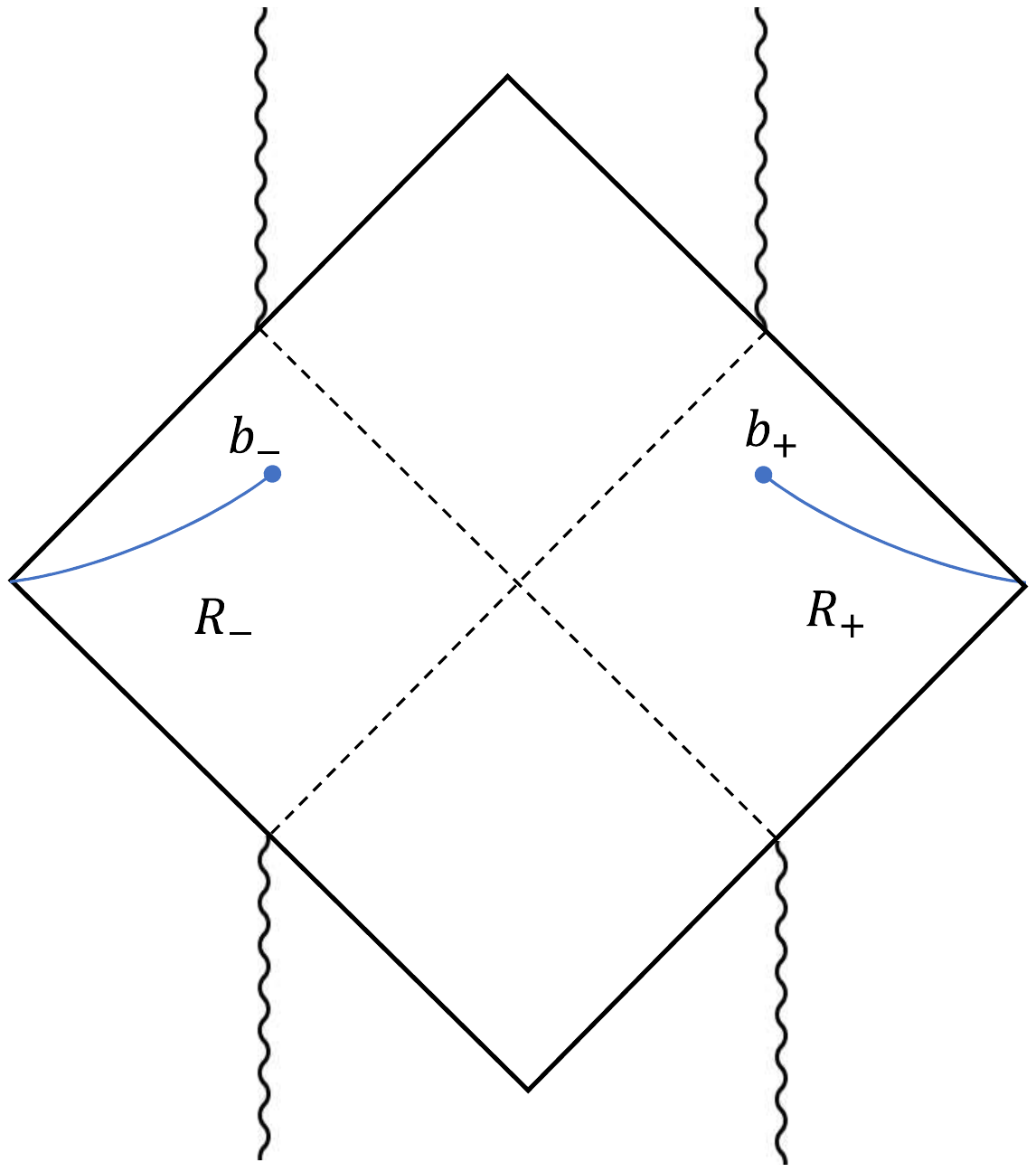}
  \caption{Penrose diagram of an eternal \\ Reissner-Nordstr{\"o}m black hole \cite{Island-RNBH} in the \\absence of island surface.}
  \label{p1}
\end{minipage}%
\begin{minipage}{.5\textwidth}
  \centering
  \includegraphics[width=0.7\linewidth]{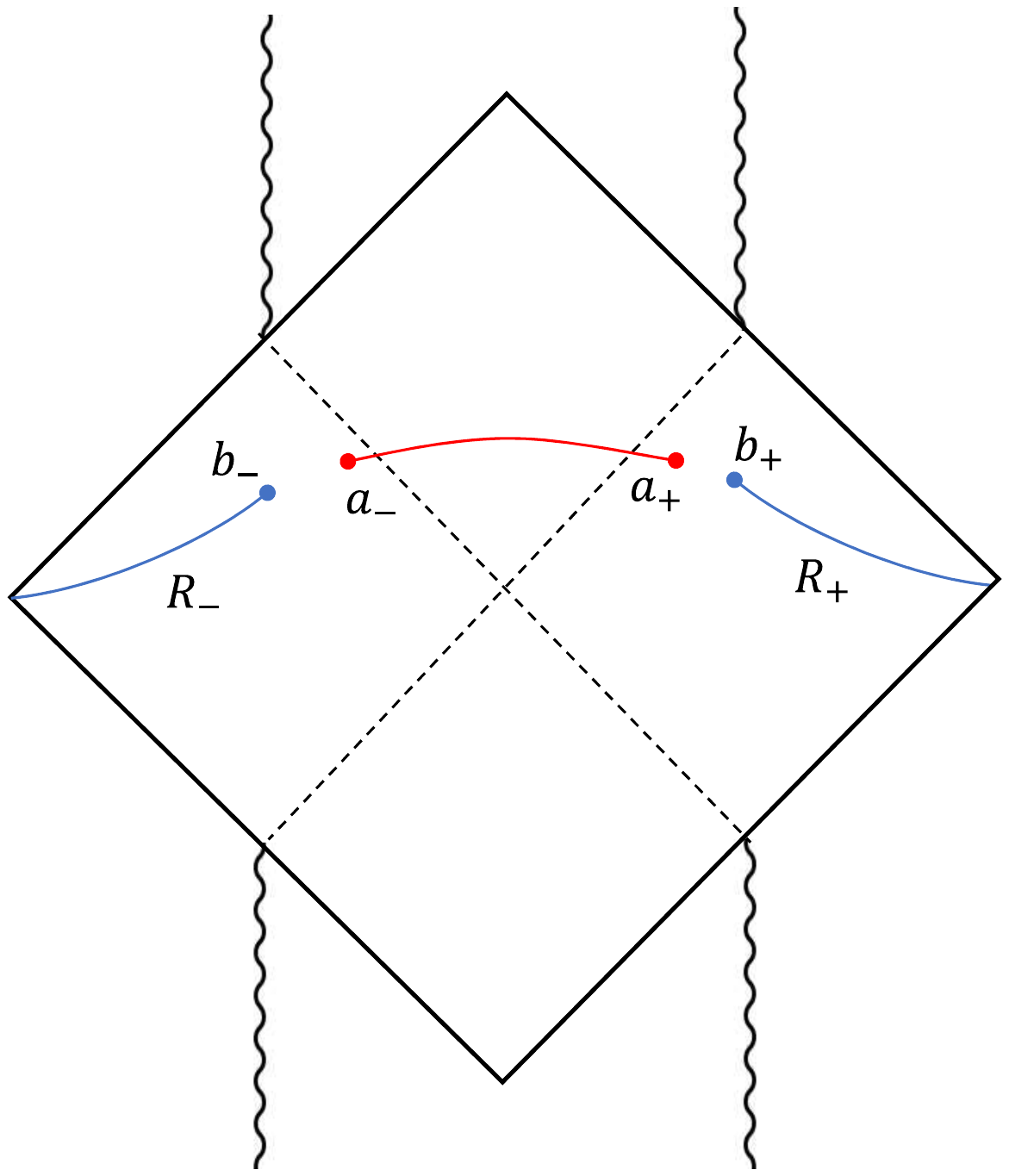}
  \caption{Penrose diagram of an eternal \\ Reissner-Nordstr{\"o}m black hole \cite{Island-RNBH} in the\\ presence of island surface.}
  \label{p2}
\end{minipage}
\end{figure}

\subsubsection{Entanglement Entropy without Island}
Initially there is no island surface therefore entanglement entropy of the Hawking radiation can be calculated using the following formula \cite{CC,CC-1,Yu-Ge}:
\begin{eqnarray}
\label{EE-formula-no-island}
S = \frac{c}{6} \log[d^2(b_+,b_-)],
\end{eqnarray}
where $b_+(t_b,b)$ and $b_-(-t_b+\iota \frac{\beta}{2},b)$ are the boundaries of the radiation regions in the right and left wedge of Reissner-Nordstr{\"o}m black hole. Formula (\ref{EE-formula-no-island}) is valid for two dimensional vacuum CFT. Since entanglement entropy formula for higher-dimensional spacetime is not known. But we can apply (\ref{EE-formula-no-island}) for higher-dimensional spacetime in s-wave approximation \cite{Island-SB} by assuming that location of the observer or cut-off surface ($b_{\pm}$) is very far away from the black hole horizon ($r_+$), i.e., $b_{\pm}\gg r_+$. Therefore in s-wave approximation we can ignore the angular part of the metric and we will be left with asymptotically flat black hole for which (\ref{EE-formula-no-island}) can be used . $\beta$ is the inverse of Hawking temperature and defined as, $\beta=2 \pi/\kappa_+$. Geodesic distance between two points $(l_1,l_2)$ is given as:
\begin{equation}
\label{d}
d(l_1,l_2)=\sqrt{g(l_1)g(l_2)(U(l_2)-U(l_1))(V(l_1)-V(l_2))}.
\end{equation}
Now using equations (\ref{U-V}), (\ref{rstar-U}), (\ref{conformal-factor}), (\ref{EE-formula-no-island}) and (\ref{d}) entanglement entropy of the Hawking radiation in the absence of island surface simplifies to \cite{Island-RNBH}:
\begin{eqnarray}
\label{EE-without-island}
S = \frac{c}{6}\log[4 g(b)^2 e^{2 \kappa_+ r_*(b)} \cosh^2 \kappa_+t],
\end{eqnarray}
when $t \rightarrow \infty$, i.e., at late times, writing $ \cosh \kappa_+t \sim e^{\kappa_+t}$, time dependent part of equation (\ref{EE-without-island}) simplifies to:
\begin{eqnarray}
\label{EE-ET-ET-1}
S_{\rm EE}^{\rm WI} \sim \frac{c}{3} \kappa_+ t .
\end{eqnarray}
Therefore from equation (\ref{EE-ET-ET-1}) we can see that entanglement entropy of the Hawking radiation in the absence of island surface is increasing linearly with time and becomes infinite at late times which leads to the information paradox for the Reissner-Nordstr{\"o}m black hole. In the next subsection we will see that at late times island appears and entanglement entropy of the Hawking radiation in the presence of island surface will be constant and dominates after the Page time. On combining the both contributions, we obtain the Page curve.

\subsubsection{Entanglement Entropy with Island}
In this subsection we are reviewing calculation of the entanglement entropy of the Hawking radiation in the presence of island surface. We can calculate it using the following formula \cite{CC,CC-1,Yu-Ge}:
\begin{eqnarray}
\label{EE-formula-island}
S({\cal R}\cup {\cal I}) = \frac{c}{3} \log\left(\frac{d(a_+,a_-)d(b_+,b_-)d(a_+,b_+)d(a_-,b_-)}{d(a_+,b_-)d(a_-,b_+)}\right),
\end{eqnarray}
where $a_+(t_a,a)$ and $a_-(-t_a+\iota \frac{\beta}{2},a)$ are the boundaries of the island surface in the right and left wedge of Reissner-Nordstr{\"o}m geometry. Entanglement entropy formula (\ref{EE-formula-island}) can be used for higher-dimensional spacetime in s-wave approximation  where we can ignore the angular part of the metric for a distant observer ($b_{\pm}\gg r_+$). At late times, $t_a,t_b \gg b >r_+$, from equations (\ref{U-V}), (\ref{rstar-U}), (\ref{conformal-factor}) and (\ref{d}), equation (\ref{EE-formula-island}) simplifies to \cite{Island-RNBH}:
\begin{eqnarray}
\label{Smatter-LT-RNBH}
& & 
S({\cal R}\cup {\cal I})^{\rm late \ times} =\frac{c}{6}\log[g^2(a)g^2(b)]+\frac{2 c}{3} \kappa_+ r_*(b)\nonumber\\
& & +\frac{c}{3}\Biggl[-2 e^{-\kappa_+(b-a)}\Biggl|\frac{a-r_+}{b-r_+}\Biggr|^{1/2}\Biggl|\frac{a-r_-}{b-r_-}\Biggr|^{-(r_-^2/r_+^2)} -e^{\kappa_+(r_*(b)-r_*(a)-2 t_b)} \Biggr].
\end{eqnarray}

Hence generalised entropy in the presence of island surface using (\ref{Island-proposal}) at late times is given by \cite{Island-RNBH}:
\begin{eqnarray}
\label{EE-LT-RNBH}
& & 
S_{gen}^{\rm late \ times}(a)=\frac{2 \pi a^2}{G_N}+\frac{c}{6}\log[g^2(a)g^2(b)]+\frac{2 c}{3} \kappa_+ r_*(b)\nonumber\\
& & +\frac{c}{3}\Biggl[-2 e^{-\kappa_+(b-a)}\Biggl|\frac{a-r_+}{b-r_+}\Biggr|^{1/2}\Biggl|\frac{a-r_-}{b-r_-}\Biggr|^{-(r_-^2/r_+^2)}- e^{\kappa_+(r_*(b)-r_*(a)-2 t_b)} \Biggr],
\end{eqnarray}
where first term in the above equation is coming from the area of the island surface. On varying equation (\ref{EE-LT-RNBH}) with respect to $a$, i.e., $\frac{\partial S_{gen}^{\rm late \ times}(a)}{\partial a}=0$, we are required to solve the following equation:
\begin{eqnarray}
\frac{4 \pi  r_+}{G_N}+\frac{c \left(\frac{r_+-r_-}{b-r_-}\right){}^{-\frac{r_-^2}{r_+^2}} e^{\kappa _+ \left(r_+-b\right)}}{3 \left(r_+-b\right) \sqrt{\frac{a-r_+}{b-r_+}}}=0,
\end{eqnarray}
solution to the above equation is,
\begin{eqnarray}
\label{a-RNBH}
a \approx r_+ + \Biggl[\frac{c^2 \left(\frac{r_+-r_-}{b-r_-}\right){}^{-\frac{2 r_-^2}{r_+^2}} G_N^2 e^{2 \kappa _+ \left(r_+-b\right)}}{144 \pi ^2 r_+^2 \left(b-r_+\right)} \Biggr],
\end{eqnarray}

Substituting the value of $a$ from equation (\ref{a-RNBH}) in equation (\ref{EE-LT-RNBH}), we obtain the total entanglement entropy of the Hawking radiation at late times as:
\begin{eqnarray}
\label{EE-RUI}
& &
S_{\rm total}^{(0)}=\frac{2\pi r_+^2}{G_N}+\frac{c}{3}  \log \left(\frac{r_- e^{2 \kappa _+ \left(b-r_+\right)} \left(\frac{r_-^2}{\left(r_+-r_-\right) \left(b-r_-\right)}\right){}^{\frac{1}{2} \left(\frac{\kappa _+}{\kappa _-}-1\right)}}{b
   \kappa _+^2}\right)+small,\nonumber\\
   & & 
   S_{\rm total}^{(0)}=2 S_{\rm BH}^{\rm (0)}+\frac{c}{3}  \log \left(\frac{r_- e^{2 \kappa _+ \left(b-r_+\right)} \left(\frac{r_-^2}{\left(r_+-r_-\right) \left(b-r_-\right)}\right){}^{\frac{1}{2} \left(\frac{\kappa _+}{\kappa _-}-1\right)}}{b
\kappa _+^2}\right)+small.
\end{eqnarray}
From the above equation we can see that total entanglement entropy is constant which dominates after the Page time and we obtain the Page curve of an eternal Reissner-Nordstr{\"o}m black hole.
\subsubsection{Page Curve}
From equation (\ref{EE-ET-ET-1}) we saw that entanglement entropy in the absence of island surface is increasing linearly with time and from equation (\ref{EE-RUI}) we saw that entanglement entropy is constant at late times. Therefore initially entanglement entropy of the Hawking radiation will increase and after the Page time island surface emerges which saturates the linear growth of entanglement entropy and reaches a constant value, which is twice of the Bekenstein-Hawking entropy of the Reissner-Nordstr{\"o}m black hole, and we obtain the Page curve of the Reissner-Nordstr{\"o}m black hole. \par
  We have plotted the Page curve of the Reissner-Nordstr{\"o}m black hole for, $M=1,Q=0.8,G_N=1$ in figure {\ref{Page-curve-RNBH}} by considering only leading order term in $G_N$ in equation (\ref{EE-RUI}). In figure {\ref{Page-curve-RNBH}}, blue line corresponds to the linear time growth of the entanglement entropy (equation (\ref{EE-ET-ET-1})) and orange line corresponds to the entanglement entropy in the presence of island surface (equation ({\ref{EE-RUI})) which dominates after the Page time and leads to the Page curve.

\begin{figure}
\begin{center}
\includegraphics[width=0.50\textwidth]{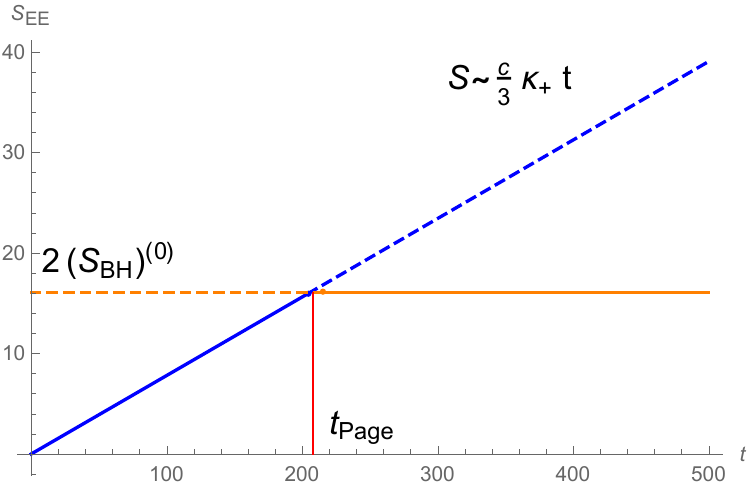}
\end{center}
\caption{Page curve of an eternal Reissner-Nordstr{\"o}m black hole by considering only leading order term in $G_N$.}
\label{Page-curve-RNBH}
\end{figure}

{\bf Page time:} Page time is defined as time at which entanglement entropy of the Hawking radiation start falling to zero for an evaporating black hole and reaches constant value for an eternal black hole. We can get the Page time by equating equations (\ref{EE-ET-ET-1}) and (\ref{EE-RUI}), which turn out to be:
\begin{eqnarray}
\label{Page-time}
t_{Page}^{(0)} \sim \frac{6 \pi  r_+^2}{c \kappa _+ G_N} =\frac{12 \pi r_+^4}{c\ G_N (r_+ - r_-)} = \frac{3 S_{\rm BH}^{(0)}}{2 \pi c T_{\rm RN}}.
\end{eqnarray}

{\bf Scrambling time:} Scrambling time is defined as time interval in which we recover the information thrown into the black hole in the form of Hawking radiation \cite{scrambling-time-1,scrambling-time-2}. It was discussed in \cite{scrambling-time-1} that for an evaporating black hole information can be quickly retrieved in the form of Hawking radiation if the black hole has evaporated half away. If the black hole has not evaporated half away then one has to wait to evaporate half away, after that information can be quickly recovered. In the language of entanglement wedge reconstruction \cite{scrambling-time-EW}, scrambling time is defined as time at which information from the cut-off surface ($r=b_+$) reaches the boundary of the island surface ($r=a_+$). When information reaches to the boundary of island surface ($r=a_+$) then entanglement entropy of Hawking radiation includes contribution from the island surface because island degrees of freedom becomes part of the entanglement wedge of the Hawking radiation and we start recovering the information thrown into the black hole in the form of Hawking radiation. If we want to send the information into the black hole from the cutoff surface ($r=b$) then time taken by information to reach the boundary of island surface ($r=a$) is given by:
\begin{eqnarray}
\label{delta-t}
t_{scr}^{(0)} =r_*(b)-r_*(a)= b-a+\frac{r_-^2}{r_+ -r_-} \log\left(\frac{a-r_-}{b-r_-}\right)-\frac{r_+^2}{r_+ -r_-} \log\left(\frac{a-r_+}{b-r_+}\right).
\end{eqnarray} 
Substituting the value of $a$ from equation (\ref{a-RNBH}) in the above equation, scrambling time of the Reissner-Nordstr{\"o}m black hole turns out to be:
\begin{eqnarray} 
\label{tscr}
t_{scr}^{(0)}\sim \frac{2 r_+^2}{(r_+ - r_-)} \log\left(\frac{\pi r_+^2}{G_N}\right) + small=\frac{1}{2 \pi T_{\rm RN}} \log\left(S_{\rm BH}^{(0)}\right) + small, 
\end{eqnarray}
where $T_{\rm RN}$ is the Hawking temperature and $S_{\rm BH}^{(0)}$ is the Bekenstein-Hawking entropy of the Reissner-Nordstr{\"o}m black hole. From equation (\ref{tscr}) we can see that scrambling time turns out to be logarithmic of black hole thermal entropy and this result is similar to \cite{scrambling-time-2} which implies that black holes are fastest scramblers.

\section{Page Curves of Charged Black Hole in HD Gravity up to ${\cal O}(R^2)$}
\label{Page-curve-R^2}
In this section we are considering general higher derivative terms at ${\cal O}(R^2)$ in the gravitational action as considered in \cite{NBH-HD} and we are going to calculate the Page curves of an eternal Reissner-Nordstr{\"o}m black hole in the presence of ${\cal O}(R^2)$ terms. In our case we have charged black hole whereas in \cite{NBH-HD} authors obtained the Page curve of  Schwarzschild black hole. Action with inclusion of ${\cal O}(R^2)$ terms and Maxwell-term is:
\begin{eqnarray}
\label{action-R^2}
S=\frac{1}{16 \pi G_N}\int d^4x \sqrt{-g}\left(R[g]+\lambda_1 R^2[g]+\lambda_2 R_{\mu \nu}[g] R^{\mu \nu}[g]+\alpha {\cal L}_{GB}[g]-F_{\mu \nu}F^{\mu \nu}\right),
\end{eqnarray}
where $R[g]$ is the Ricci scalar, $R_{\mu\nu}[g]$ is the Ricci tensor and ${\cal L}_{GB}$ is the Gauss-Bonnet term, ${\cal L}_{GB}=R_{\mu\nu\rho\sigma}[g]R^{\mu\nu\rho\sigma}[g]-4 R_{\mu\nu}[g]R^{\mu\nu}[g]+R^2[g]$ in four dimensions. Wald entropy of the four dimensional charged black hole in higher derivative gravity can be calculated using the formula given in \cite{Wald-Entropy,Wald-Entropy-2} and used in \cite{McTEQ} to calculate Wald entropy in the presence of ${\cal O}(R^4)$ terms in ${\cal M}$-theory action \cite{HD-Roorkee}. The formula is:
\begin{eqnarray}
\label{Wald-Entropy-defn}
S_{\rm Wald}=\frac{1}{4 G_N}\oint \sqrt{h} \left(\frac{\partial {\cal L}}{\partial R_{\mu\nu\rho\sigma}}\right)\epsilon_{\mu\nu}\epsilon_{\rho\sigma} d^2x,
\end{eqnarray}
where $h$ is the determinant of the induced metric on co-dimension two surface. Writing the metric ({\ref{metric-RNBH-original}) in the following form:
\begin{equation}
\label{metric-redefined}
ds^2=-\left(\frac{(r-r_+)(r-r_-)}{r^2}\right)dt^2+\frac{dr^2}{\left(\frac{(r-r_+)(r-r_-)}{r^2}\right)}+r^2\left(d\theta^2+\sin^2\theta d\phi^2\right),
\end{equation}

From the action (\ref{action-R^2}) we find that:
\begin{eqnarray}
\label{der-Lag}
\frac{\partial {\cal L}}{\partial R_{\mu\nu\rho\sigma}}=\Biggr[g^{\nu\sigma}g^{\mu\rho}(1+2\alpha R[g])+2\alpha R^{\mu\nu\rho\sigma}[g]-8\alpha g^{\mu\rho}R^{\nu\sigma}[g]\Biggl]+2 \lambda_1 R[g] g^{\nu\sigma}g^{\mu\rho}+2\lambda_2 g^{\mu\rho}R^{\nu\sigma}[g],\nonumber\\
\end{eqnarray}
where the terms written in the square bracket will also be used in calculation of Wald entropy of charged black hole in Einstein-Gauss-Bonnet gravity in section {\ref{Page-curves-CEGB-BH}}. For the metric (\ref{metric-redefined}), Wald entropy corresponding to the action (\ref{action-R^2}) using equation (\ref{der-Lag}) turns out to be:
\begin{eqnarray}
\label{Wald-Entropy-R^2-gravity}
& &
S_{\rm Wald}^{(R^2)}=S_{\rm BH}^{(\alpha)}=\frac{1}{4 G_N}\oint \sqrt{h} \left(\frac{\partial {\cal L}}{\partial R_{trtr}}\right) d\theta d\phi= \frac{\pi}{G_N}\left(r_+^2+\frac{4 \alpha(r_+ +6 r_-)}{r_+}-2\lambda_2\left(\frac{r_- r_+}{r_+^2}\right)\right).\nonumber\\
\end{eqnarray}
Since  $r_+ \gg r_-$, therefore above equation simplifies to: 
\begin{eqnarray}
\label{Wald-Entropy-R^2-gravity-simp}
& &
S_{\rm Wald}^{(R^2)}=S_{\rm BH}^{(\alpha)}\sim \frac{\pi}{G_N}\left(r_+^2+4 \alpha\right).
\end{eqnarray}
In general higher derivative gravity theories holographic entanglement entropy can be calculated using the following formula \cite{Dong}:
\begin{eqnarray}
\label{EE-HD-Formula}
& &
\hskip -0.3in S_{\rm gravity}= \frac{1}{4 G_N} \int d^2 y \sqrt{h} \Biggl[ -\frac{\partial {\cal L} }{\partial R_{{\mu_1} {\nu_1} {\rho_1} {\sigma_1}}} \epsilon_{{\mu_1} {\nu_1}} \epsilon_{{\rho_1} {\sigma_1}}+\sum_\alpha \left(\frac{\partial^2{\cal L} }{\partial R_{{\mu_3} {\nu_3} {\rho_3} {\sigma_3}} \partial R_{{\mu_1} {\nu_1} {\rho_1} {\sigma_1}}}\right)_\alpha \frac{2 K_{ {\lambda_3} {\rho_3} {\sigma_3}}K_{{\lambda_1} {\rho_1} {\sigma_1}}}{q_{\alpha}+1} \nonumber\\
& & \times [\left(n_{{\mu_3}{\mu_1}}n_{{\nu_3}{\nu_1}} -\epsilon_{{\mu_3}{\mu_1}}\epsilon_{{\nu_3}{\nu_1}} \right)n^{{\lambda_3}{\lambda_1}} +\left(n_{{\mu_3}{\mu_1}}\epsilon_{{\nu_3}{\nu_1}} +\epsilon_{{\mu_3}{\mu_1}}n_{{\nu_3}{\nu_1}} \right)\epsilon^{{\lambda_3}{\lambda_1}}]\Biggr],
\end{eqnarray}
where
\begin{eqnarray}
& &
n_{\mu \nu}=n_\mu^{(i)}n_\nu^{(j)}g_{ij}, \nonumber\\
& & \epsilon_{\mu \nu}=n_\mu^{(i)}n_\nu^{(i)}\epsilon_{ij},\nonumber\\
& & \epsilon_{\mu \nu}\epsilon_{\rho \sigma}=n_{\mu \rho}n_{\nu \sigma}-n_{\mu \sigma}n_{\nu \rho}, \nonumber\\
& & K_{\lambda \mu \nu}=n_\lambda^{(i)}m_\mu^{(a)}n_\nu^{(b)} K_{iab},
\end{eqnarray}
wher $m_\mu^{(a)}$ are unit orthogonal vectors along $y^a$ (tangential) directions, $n_\mu^{(i)}$ are the untit normal vectors along the normal directions, $K_{\lambda \mu \nu}$ is the extrinsic curvature and $q_\alpha$ in equation (\ref{EE-HD-Formula}) is a number which is sum of the certain combination of Riemann tensor components and extrinsic curvatures which can be obtained by first calculating the second order derivative of Lagrangian and labelling each term by $\alpha$ and then performing expansion of Riemann tensor components along certain directions 
(for more details, see \cite{Dong}). We can write the total entanglement entropy in the presence of higher derivative terms as:
\begin{eqnarray}
\label{S-total-defn}
S_{\rm total}=S_{\rm gravity}+S_{\rm matter}(R \cup {\cal I}).
\end{eqnarray}
 For the action (\ref{action-R^2}), gravity contribution to the entanglement entropy turns out to be \cite{NBH-HD}:
\begin{eqnarray}
\label{EE-gravity-R^2}
S_{\rm gravity} = \frac{A[\partial {\cal I}]}{4 G_{N, \rm ren}}+\frac{1}{4 G_{N, \rm ren}} \int_{\partial  {\cal I}} \left(2 \lambda_{1, \rm ren} R[g]+ \lambda_{2, \rm ren} \sum_{i=1}^2[R_{\mu\nu}[g]n^{\mu}_in^\nu_i-\frac{1}{2}K_i K_i]+2 \alpha_{\rm ren} R[\partial  {\cal I}]\right),\nonumber\\
\end{eqnarray}
where $i$ corresponds to the normal directions, $K_i$ being trace of extrinsic curvature, $K_{i,\mu\nu}=-h^\alpha_\mu h_{\nu\beta}$, $h_{\mu\nu}$ is the induced metric on boundary of the island surface. Further, $G_{N, \rm ren}, \lambda_{1, \rm ren},\lambda_{2, \rm ren}$ and $\alpha_{\rm ren}$ are the renormalised Newton constant and coupling constants appearing in higher derivative gravity action (\ref{action-R^2}). These are used to absorb the UV divergences of the von Neumann entropy of matter field discussed in \cite{NBH-HD}. Matter contribution to the entanglement entropy of the Hawking radiation in the presence of island surface can be obtained using equations (\ref{U-V}), (\ref{rstar-U}), (\ref{conformal-factor}), (\ref{d}) and (\ref{EE-formula-island}) and is  given below \cite{Island-RNBH}{\footnote{Matter contribution to the entanglement entropy, $S_{\rm matter}(R\cup {\cal I})$, up to some extent is same as \cite{Island-RNBH}. Therefore we are writing the final result up to that point.}:
\begin{eqnarray}
& & 
S_{\rm matter}(R \cup {\cal I})=\frac{c}{6}\log[2^4 g^2(a)g^2(b)\cosh^2(\kappa_+ t_a)\cosh^2(\kappa_+ t_b)]+\frac{c}{3} \kappa_+(r_*(a)+r_*(b))\nonumber\\
& & +\frac{c}{3}\Biggl[\frac{\cosh\left(\kappa_+(r_*(a)-r_*(b)) \right)-\cosh\left(\kappa_+(t_a-t_b)\right)}{\cosh\left(\kappa_+(r_*(a)-r_*(b)) \right)+\cosh\left(\kappa_+(t_a-t_b) \right)} \Biggr].
\end{eqnarray}  
At late times, $t_a,t_b \gg b > r_+$, above equation simplifies to the following form:
\begin{eqnarray}
\label{EE-matter-late-times-R^2}
& & 
S_{\rm matter}^{\rm late \ time}(R \cup {\cal I})=\frac{c}{6}\log[g^2(a)g^2(b)]+\frac{2 c}{3} \kappa_+ r_*(b)\nonumber\\
& & +\frac{c}{3} \Biggl[-2 e^{-\kappa_+(b-a)} \Biggl|\frac{a-r_+}{b-r_+}\Biggr|^{1/2}\Biggl|\frac{a-r_-}{b-r_-}\Biggr|^{-(r_-^2/r_+^2)}-e^{\kappa_+\left(r_*(b)-r_*(a)-2t_b\right)} \Biggr].
\end{eqnarray}
Therefore generalised entropy for the action (\ref{action-R^2}) is \cite{NBH-HD}:
{\footnotesize
\begin{eqnarray}
\label{gen-entropy-HD-gravity}
& & 
S_{gen}(r)={\rm Min}_{\cal I} \Biggl[{\rm Ext}_{\cal I}\Biggl(\frac{A[\partial {\cal I}]}{4 G_{N, \rm ren}}+\frac{1}{4 G_{N, \rm ren}} \int_{\partial  {\cal I}} \left(2 \lambda_{1, \rm ren} R[g]+ \lambda_{2, \rm ren} \sum_{i=1}^2[R_{\mu\nu}[g]n^{\mu}_in^\nu_i-\frac{1}{2}K_i K_i]+2 \alpha_{\rm ren} R[\partial  {\cal I}]\right)  
 \nonumber\\
 & & \hskip 1in + S_{\rm matter}(R\cup {\cal I})\Biggr)
\Biggr].\nonumber\\
\end{eqnarray}
}
For the metric (\ref{metric-RNBH-original}), we find that:
$R[g]=0, R[\partial  {\cal I}]=\frac{2}{a^2}, n_1^{t}=\frac{1}{\sqrt{F(r)}},n_2^{r}=\sqrt{F(r)}$,$K_1=0$ and $K_2=-\frac{2}{r}\sqrt{F(r)}$.
Therefore, 
\begin{eqnarray}
& & 
\sum_{i=1}^2 R_{\mu\nu}[g]n^{\mu}_in^\nu_i =0,\nonumber\\
& & 
K_1 K_1=0, \nonumber\\
& & K_2 K_2=\frac{4}{a^2}\left(1-\frac{2 M}{a}+\frac{Q^2}{a^2}\right).
\end{eqnarray}
Hence gravitational contribution to the entanglement entropy of the Hawking radiation (\ref{EE-gravity-R^2}) simplifies to,
\begin{eqnarray}
\label{EE-gravity-simp-R^2}
& &
S_{\rm gravity}(a)=\frac{2 \pi}{G_{N,\rm ren}}\Biggl[a^2-2 \lambda_{2,\rm ren} \left(1-\frac{2 M}{a}+\frac{Q^2}{a^2}\right)+4 \alpha_{\rm ren}\Biggr],
\end{eqnarray}

Therefore total generalised entropy will be given by the sum of entanglement entropy contributions from gravity part (\ref{EE-gravity-simp-R^2}) and matter (\ref{EE-matter-late-times-R^2}) part both as:
\begin{eqnarray}
\label{generalised-EE-R^2}
& & 
S_{\rm gen}(a)=S_{\rm gravity}(a)+S_{matter}^{\rm late \ times}(R \cup {\cal I})(a) \nonumber\\
& & S_{\rm gen}(a) \sim \frac{2 \pi}{G_{N,\rm ren}}\Biggl[a^2-2 \lambda_{2,\rm ren} \left(1-\frac{2 M}{a}+\frac{Q^2}{a^2}\right)+4 \alpha_{\rm ren}\Biggr]+\frac{c}{6}\log[g^2(a)g^2(b)]+\frac{2 c}{3} \kappa_+ r_*(b)\nonumber\\
& & \hskip 0.75in  +\frac{c}{3} \Biggl[-2 e^{-\kappa_+(b-a)} \Biggl|\frac{a-r_+}{b-r_+}\Biggr|^{1/2}\Biggl|\frac{a-r_-}{b-r_-}\Biggr|^{-(r_-^2/r_+^2)}\Biggr],
\end{eqnarray}
where $r_*(b),g(a)$ and $g(b)$ can be substituted from equations (\ref{rstar-U}) and (\ref{conformal-factor}) in equation (\ref{generalised-EE-R^2}). $t_b$ dependent term in equation (\ref{EE-matter-late-times-R^2}) is small and hence has been ignored. Location of the island surface can be found by extremizing the above equation with respect to $a$, i.e.
\begin{eqnarray}
\frac{\partial S_{\rm gen}(a) }{\partial a} \sim \frac{4 \pi  \left(-2 M r_+ \lambda _{2,\rm ren}+2 Q^2 \lambda _{2,\rm ren}+r_+^4\right)}{r_+^3 G_N}+\frac{c \left(\frac{r_+-r_-}{b-r_-}\right){}^{-\frac{r_-^2}{r_+^2}} e^{\kappa _+
   \left(r_+-b\right)}}{3 \left(r_+-b\right) \sqrt{\frac{a-r_+}{b-r_+}}}=0,
\end{eqnarray}
solution to the above equation is,
\begin{eqnarray}
\label{a-R^2-gravity}
& & 
a \approx r_+ +\frac{c^2 r_+^6 \left(\frac{r_+-r_-}{b-r_-}\right){}^{-\frac{2 r_-^2}{r_+^2}} G_N^2 e^{2 \kappa _+ \left(r_+-b\right)}}{144 \pi ^2 \left(b-r_+\right) \left(2 \lambda _{2,\rm ren} \left(Q^2-M r_+\right)+r_+^4\right){}^2}.
\end{eqnarray}
From the above equation we can see that island lies outside the black hole horizon and this leads to causality paradox. It was shown in \cite{island-o-h} that this result appears in all the two sided eternal black holes or black holes in Hartle-Hawking state and one can restore the causality by quantum focusing conjecture (QFC) \cite{QFC}. Idea of  \cite{island-o-h} is that when we decouple the black hole from the bath then finite amount of energy flux will be produced and this energy flux pushes the black hole horizon outwards and therefore island always lies behind the horizon. It was discussed in \cite{island-coupled} that finite amount of energy will also be produced even when we couple the black hole to bath and this energy flux pushes the horizon outwards which implies that island lies behind the horizon similar to the decoupling process and we can get rid of causality paradox.

Substituting value of $a$ from equation (\ref{a-R^2-gravity}) in equation (\ref{generalised-EE-R^2}), generalised entropy  simplifies to the following form\footnote{
\label{Smatter-footnote}In our case exponential factor in the ${\cal O}(G_N^0)$ term in the total entanglement entropies (\ref{S-total-HD}) and (\ref{total-EE-EGB-simp}) is different from \cite{Island-RNBH}. Which can be seen as follow:
Relevant term is $S \sim \frac{c}{3} \log[g(a)g(b)]$. Now using equation (\ref{conformal-factor}). This term can be written as:
\begin{eqnarray}
S_{\rm matter} \sim \frac{c}{3} \log\left(\frac{r_+ r_-}{ab \kappa_+^2}\left(\frac{r_-^2}{(a-r_-)(b-r_-)}\right)^{\frac{1}{2}\left(\frac{\kappa_+}{\kappa_-}-1\right)} e^{-\kappa_+(a+b)}\right).
\end{eqnarray}
Now substituting $a\approx r_+$ in the above equation we obtain:
\begin{eqnarray}
S_{\rm matter} \sim \frac{c}{3}  \log \left(\frac{r_- e^{- \kappa _+ \left(b+r_+\right)} \left(\frac{r_-^2}{\left(r_+-r_-\right) \left(b-r_-\right)}\right){}^{\frac{1}{2} \left(\frac{\kappa _+}{\kappa _-}-1\right)}}{b
   \kappa _+^2}\right).
\end{eqnarray}
This will not affect the Page curve of the Reissner-Nordstr{\"o}m black hole in the $\alpha \rightarrow 0$ limit because to obtain the Page curve we are considering only leading order term in equations (\ref{S-total-HD}) and (\ref{total-EE-EGB-simp}) which in the $\alpha \rightarrow 0$ limit reduces to leading order term in equation (\ref{EE-RUI}).
Same thing can also be verified by substituting the value of $a$ from equation (\ref{a-R^2-gravity}) in equation (\ref{generalised-EE-R^2}) and performing small $G_N$ expansion and retaining the terms up to ${\cal O}(G_N^0)$.
}:
\begin{eqnarray}
\label{S-total-HD}
& &
S_{\rm total}^{{\cal O}(R^2)}=S_{\rm gravity}+S_{\rm matter}({\cal R}\cup {\cal I}), \nonumber\\
& & S_{\rm total}^{{\cal O}(R^2)} \sim \frac{2 \pi  \left( r_+^2+4 \alpha_{\rm ren}\right)}{G_N}+\frac{c}{3}  \log \left(\frac{r_- e^{- \kappa _+ \left(b+r_+\right)} \left(\frac{r_-^2}{\left(r_+-r_-\right) \left(b-r_-\right)}\right){}^{\frac{1}{2} \left(\frac{\kappa _+}{\kappa _-}-1\right)}}{b
   \kappa _+^2}\right)+ {\cal O}(G_N), \nonumber\\
   & &S_{\rm total}^{{\cal O}(R^2)} =2 S_{\rm BH}^{(\alpha)}+\frac{c}{3}  \log \left(\frac{r_- e^{- \kappa _+ \left(b+r_+\right)} \left(\frac{r_-^2}{\left(r_+-r_-\right) \left(b-r_-\right)}\right){}^{\frac{1}{2} \left(\frac{\kappa _+}{\kappa _-}-1\right)}}{b \kappa _+^2}\right)+ {\cal O}(G_N).
\end{eqnarray}
Therefore total entanglement entropy in the presence of island surface is twice of the Bekenstein-Hawking entropy of the black hole plus matter contribution. If we consider only leading order term in $G_N$ then we can see that total entanglement entropy of an eternal Reissner-Nordstr{\"o}m black hole in the presence of ${\cal O}(R^2)$ terms reaches a constant value consistent with the literatures. It is interesting to notice that $\lambda_{2,\rm ren}$ dependence is appearing in higher order in $G_N$ similar to \cite{NBH-HD}. In the $\alpha \rightarrow 0$ then equation (\ref{S-total-HD}) reduces to,
\begin{equation}
S_{\rm total}^{(0)}=\frac{2 \pi r_+^2}{G_N}+\frac{c}{3}  \log \left(\frac{r_- e^{- \kappa _+ \left(b+r_+\right)} \left(\frac{r_-^2}{\left(r_+-r_-\right) \left(b-r_-\right)}\right){}^{\frac{1}{2} \left(\frac{\kappa _+}{\kappa _-}-1\right)}}{b
   \kappa _+^2}\right). 
\end{equation}
which is the almost same as in \cite{Island-RNBH}. From equation (\ref{S-total-HD}), it is clear that total entanglement entropy at late times is constant. Combining equations (\ref{EE-ET-ET-1}) and (\ref{S-total-HD}) we obtain the Page curves. Considering only leading order term in equation (\ref{S-total-HD}), and substituting $r_+=M+\sqrt{M^2-Q^2}$,  we obtain:
\begin{eqnarray}
\label{S-total-HD-simp}
& &
S_{\rm total}= \frac{2 \pi  \left( \left(M+\sqrt{M^2-Q^2}\right)^2+4 \alpha\right)}{G_N},
\end{eqnarray}
where $\alpha$ in the above equation (and anywhere in this section) is $\alpha_{\rm ren}$, for the sake of simplicity we have just written $\alpha$. We have plotted the Page curves of the charged black hole in the presence of ${\cal O}(R^2)$ terms for, $M=1,Q=0.8,G_N=1$, for various values of the Gauss-Bonnet coupling ($\alpha$) in figure {\ref{SEE-versus-alpha}} .\par
In figure {\ref{SEE-versus-alpha}}, diagonal green line corresponds to the linear time growth of the entanglement entropy of the Hawking radiation (\ref{EE-ET-ET-1}) and red, green and blue horizontal lines which are constant in time correspond to the leading order term in $G_N$ in entanglement entropy in the presence of island surface (\ref{S-total-HD-simp}) for $\alpha=0.2,0,-0.2$. From the graph it is clear that as $\alpha$ is increasing Page curves are shifting towards later time and when  $\alpha$ is decreasing Page curves are shifting towards earlier time. $t_{1_{\rm Page}},t_{2_{\rm Page}}$ and $t_{3_{\rm Page}}$ in the figure \ref{SEE-versus-alpha} are Page time for $\alpha=0.2,0,-0.2$ respectively.
\begin{figure}
\begin{center}
\includegraphics[width=0.70\textwidth]{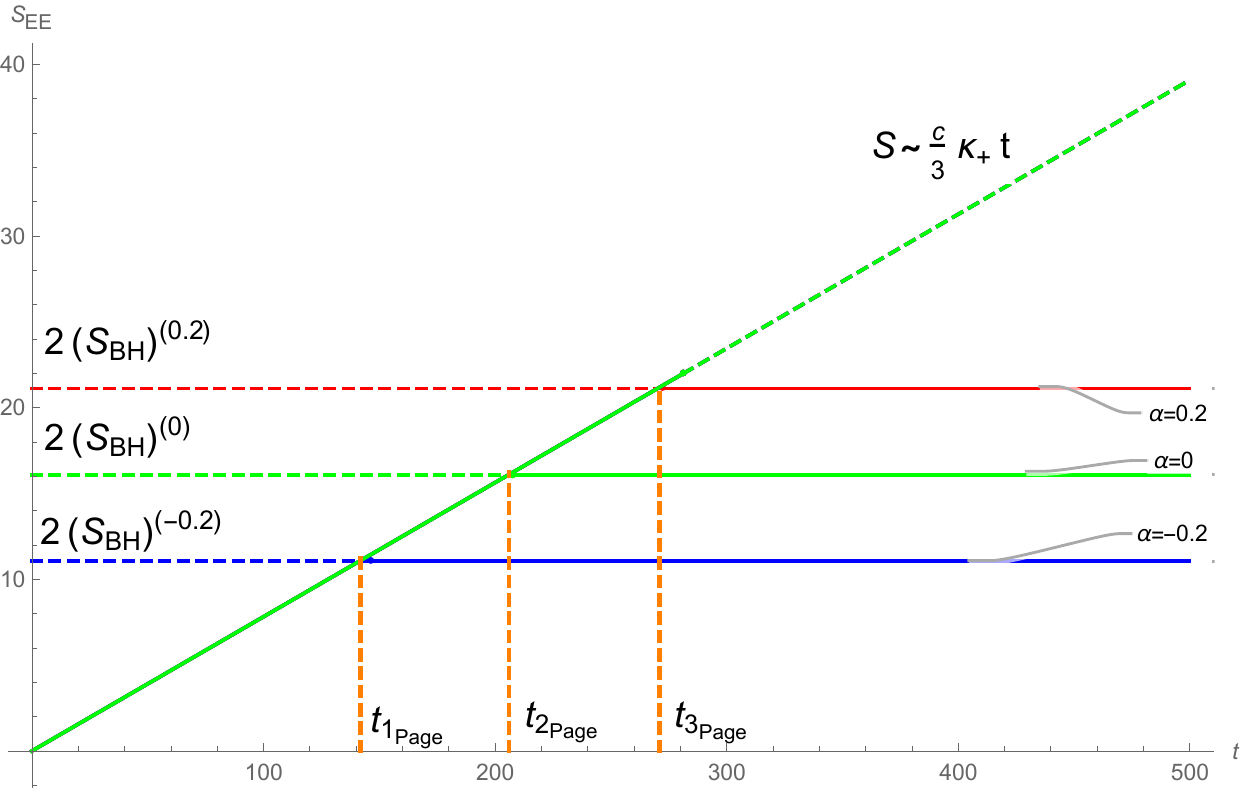}
\end{center}
\caption{Page curves of an eternal Reissner-Nordstr{\"o}m black hole for various values of Gauss-Bonnet coupling($\alpha$) by considering only leading order term in $G_N$ in total entanglement entropy in the presence of island surface at late times.}
\label{SEE-versus-alpha}
\end{figure}
\par
{\bf Page Time:} Since at $t=t_{\rm Page}^{\rm HD}$, $S_{\rm EE}^{\rm WI}=S_{\rm total}$, therefore we obtain the Page time in the presence of ${\cal O}(R^2)$ terms by equating equations (\ref{EE-ET-ET-1}) and (\ref{S-total-HD}) and is given below:
\begin{eqnarray}
\label{Page-time-HD}
& & 
t_{\rm Page}^{\rm HD} \sim  \frac{6 \pi  \left(r_+^2+4 \alpha\right)}{c \kappa _+ G_N}=\frac{3 S_{\rm BH}^{(0)}}{2 \pi c T_{\rm RN}}+\frac{12 \alpha}{c \ G_N T_{\rm RN}}=t_{\rm Page}^{(0)}+\frac{12 \alpha}{c \ G_N T_{\rm RN}}.
\end{eqnarray}
From the above equation it is clear that in $\alpha \rightarrow 0$ limit, the Page time of the Reissner-Nordstr{\"o}m black hole in the presence of ${\cal O}(R^2)$ terms reduces to Page time of the Reissner-Nordstr{\"o}m black hole (\ref{Page-time}) without higher derivative terms.
\begin{eqnarray}
t_{\rm Page}^{(0)} \sim \frac{6 \pi r_+^2}{c \kappa _+ G_N}.
\end{eqnarray}
Since in equations (\ref{S-total-HD}) and (\ref{Page-time-HD}) , Gauss-Bonnet coupling ($\alpha$) is coming with positive sign, therefore we obtain the Page curves at later or earlier time when $\alpha$ is increasing or decreasing.
\par
{\bf Scrambling time:} Substituting the value of $a$ from equation (\ref{a-R^2-gravity}) in equation (\ref{delta-t}), scrambling time of the charged black hole in the presence of ${\cal O}(R^2)$ terms turns out to be:
\begin{eqnarray} 
\label{tscr-HD}
t_{scr}^{\rm HD}\sim \frac{2 r_+^2}{r_+-r_-}{ \log \left(\frac{\pi \left(2 \lambda _{2,\rm ren} \left(Q^2-M r_+\right)+r_+^4\right)}{r_+^2 G_N}\right)}+ small .
\end{eqnarray}
In the above equation if we take $\lambda _{2,\rm ren} \rightarrow 0$ limit then we obtain the scrambling time of the Reissner-Nordstr{\"o}m black hole (\ref{tscr}) in the absence of higher derivative terms. Performing small $\lambda _{2,\rm ren}$ expansion of the equation (\ref{tscr-HD}), we obtain  
\begin{eqnarray}
\label{tscr-small-lambda}
& &
t_{\rm scr}^{\rm HD} \sim \frac{2 r_+^2}{r_+-r_-} {\log \left(\frac{\pi r_+^2}{G_N}\right)}+\frac{2 \lambda _{2,\rm ren} \left(Q^2-M r_+\right)}{\left(r_+-r_-\right) r_+^2},\nonumber\\
& & =\frac{1}{2 \pi T_{\rm RN}}  {\log \left(S_{\rm BH}^{(0)}\right)}+\frac{2 \lambda _{2,\rm ren} \left(Q^2-M r_+\right)}{\left(r_+-r_-\right) r_+^2}=t_{\rm scr}^{(0)}+\frac{2 \lambda _{2,\rm ren} \left(Q^2-M r_+\right)}{\left(r_+-r_-\right) r_+^2}.
\end{eqnarray}
From the above equation it is clear that scrambling time of the charged black hole in the presence of higher derivative terms will increase or decrease depending on whether second term in equation (\ref{tscr-small-lambda}) is positive or negative.

\section{Page Curves of Charged Black Hole in Einstein-Gauss-Bonnet Gravity}
\label{Page-curves-CEGB-BH}
In this section we are going to calculate the Page curves of charged black hole in Einstein-Gauss-Bonnet gravity using Dong formula (\ref{EE-HD-Formula}). Working action with vanishing cosmological constant is \cite{Charged-GB-BH}:
\begin{eqnarray}
\label{action-EGB}
S=\frac{1}{2 \kappa^2}\int d^4x \sqrt{-g}[R[g]+\alpha {\cal L}_{\rm GB}-F_{\mu \nu}F^{\mu \nu}],
\end{eqnarray}
where $R[g]$ is the Ricci scalar, ${\cal L}_{\rm GB}$ is the Gauss-Bonnet term (\ref{LGB-F}) and $F_{\mu \nu}$ is the field strength tensor corresponding to gauge field $A_\mu$ in four dimensions. Wald entropy of the charged black hole in Einstein-Gauss-Bonnet gravity can be obtained from equations (\ref{Wald-Entropy-defn}),(\ref{metric-redefined}),(\ref{der-Lag}) and (\ref{action-EGB}) and is given as:
\begin{eqnarray}
\label{Wald-Entropy-EGB}
S_{\rm Wald}^{(\rm EGB)}=S_{\rm BH}^{(\alpha)}= \frac{\pi}{G_N}\left(r_+^2+\frac{4 \alpha(r_+ +6 r_-)}{r_+}\right)
\sim  \frac{\pi}{G_N}\left(r_+^2+4 \alpha\right).
\end{eqnarray}  
Holographic entanglement entropy for the action (\ref{action-EGB}) was calculated in \cite{BSS,Dong,BS} and is given below:
{\footnotesize
\begin{eqnarray}
\label{EE-GB}
& &
 S_{\rm gravity}= \frac{1}{4 G_N} \int d^2y \sqrt{h} \Biggl[1+ \alpha \Biggl( 2 R[g] - 4 \left( R_{\mu\nu}[g]n^{\nu i} n^{\mu}_i -\frac{1}{2}K^{i}K_{i}\right) +2 \left(R_{\mu\nu\rho\sigma}[g]n^{\mu i} n^{\nu j} n^{\rho}_i n^{\sigma}_j - K^{i}_{ab}K_{i}^{ab} \right)\Biggr) \Biggr], \nonumber\\
\end{eqnarray}
}
where $h$ is the determinant of the following induced metric on constant $t,r=a$ surface:
\begin{eqnarray}
\label{induced-metric}
ds^2=h_{ab}dx^adx^b=a^2\left(d\theta^2+ \sin^2\theta d\phi^2\right).
\end{eqnarray}Now we are going to simplify equation (\ref{EE-GB}) using Gauss-Codazzi equation \cite{NBH-HD,BS},
\begin{eqnarray}
\label{GC-equation}
R[g]=R[\partial {\cal I}]-R_{\mu\nu\rho\sigma}[g]n^{\mu i} n^{\nu j}n^{\rho}_i n^{\sigma}_j+ 2 R_{\mu\nu}[g]n^{\nu i} n^{\mu}_i  - K^i K_i+K^i_{ab}K_i^{ab}.
\end{eqnarray}
From equations (\ref{EE-GB}) and (\ref{GC-equation}) entanglement entropy in Einstein-Gauss-Bonnet gravity simplifies to the following form (which is the gravitational contribution the generalised entropy):
\begin{eqnarray}
\label{EE-EGB-gravity}
S_{\rm gravity}= \frac{1}{4 G_N}  \int d^2y \sqrt{h} \left(1+ 2 \alpha R[\partial {\cal I}] \right),
\end{eqnarray}
and matter contribution to the entanglement entropy is \cite{Island-RNBH}{\footnote{Matter contributions to the entanglement entropy is same as \cite{Island-RNBH} up to some point which can be obtained using equations (\ref{U-V}), (\ref{rstar-U}), (\ref{conformal-factor}), (\ref{d}) and (\ref{EE-formula-island}). Therefore we are writing the final result up to that point. }:
\begin{eqnarray}
& & 
S_{\rm matter}(R \cup {\cal I})=\frac{c}{6}\log[2^4 g^2(a)g^2(b)\cosh^2(\kappa_+ t_a)\cosh^2(\kappa_+ t_b)]+\frac{c}{3} \kappa_+(r_*(a)+r_*(b))\nonumber\\
& & +\frac{c}{3}\Biggl[\frac{\cosh\left(\kappa_+(r_*(a)-r_*(b)) \right)-\cosh\left(\kappa_+(t_a-t_b)\right)}{\cosh\left(\kappa_+(r_*(a)-r_*(b)) \right)+\cosh\left(\kappa_+(t_a-t_b) \right)} \Biggr]
\end{eqnarray}  
Entanglement entropy at late times, $t_a,t_b \gg b > r_+$ is given by the following expression:
\begin{eqnarray}
\label{EE-matter-late-times}
& & 
S_{\rm matter}^{\rm late \ time}(R \cup {\cal I})=\frac{c}{6}\log[g^2(a)g^2(b)]+\frac{2 c}{3} \kappa_+ r_*(b)\nonumber\\
& & +\frac{c}{3} \Biggl[-2 e^{-\kappa_+(b-a)} \Biggl|\frac{a-r_+}{b-r_+}\Biggr|^{1/2}\Biggl|\frac{a-r_-}{b-r_-}\Biggr|^{-(r_-^2/r_+^2)}-e^{\kappa_+\left(r_*(b)-r_*(a)-2t_b\right)} \Biggr],
\end{eqnarray}
where $r_*(b)$ and $g(a),g(b)$ can be obtained from equations (\ref{rstar-U}) and (\ref{conformal-factor}).
From equations (\ref{EE-EGB-gravity}) and (\ref{EE-matter-late-times}), generalised entropy of the charged black hole in the Einstein-Gauss-Bonnet gravity is: 
{\footnotesize
\begin{eqnarray}
\label{gen-entropy-EGB-gravity}
& & 
\hskip -1in S_{gen}(a)={\rm Min}_{\cal I} \Biggl[{\rm Ext}_{\cal I}\Biggl(\frac{A[\partial {\cal I}]}{4 G_{N}}+\frac{1}{4 G_{N}} \int d^2y \sqrt{h} \left( 2 \alpha  R[\partial {\cal I}] \right)
 + S_{\rm matter}({\cal R} \cup {\cal I}) \Biggr)
\Biggr], \nonumber\\
\end{eqnarray}
}
where first term in the above equation is coming from the first term in equation (\ref{EE-EGB-gravity}). For the induce metric (\ref{induced-metric}), we find that:
$ R[\partial  {\cal I}]=\frac{2}{a^2}$. Therefore gravitational contribution to entanglement entropy (\ref{EE-EGB-gravity}) simplifies to:
\begin{eqnarray}
\label{EE-gravity-simp}
S_{\rm gravity}=\frac{2 \pi  a^2}{G_N}\left(1+\frac{4 \alpha}{a^2}
\right),
\end{eqnarray}
 Now generalised entropy will be given by sum of equations (\ref{EE-matter-late-times}) and (\ref{EE-gravity-simp}),
\begin{eqnarray}
\label{total-EE-EGB}
& &  S_{\rm gen}(a) \sim \frac{2 \pi  a^2}{G_N}\left(1+\frac{4 \alpha}{a^2}
\right)+  \frac{c}{6}\log[g^2(a)g^2(b)]+\frac{2 c}{3} \kappa_+ r_*(b) \nonumber\\
& &+\frac{c}{3} \Biggl[-2 e^{-\kappa_+(b-a)} \Biggl|\frac{a-r_+}{b-r_+}\Biggr|^{1/2}\Biggl|\frac{a-r_-}{b-r_-}\Biggr|^{-(r_-^2/r_+^2)} \Biggr], \nonumber\\
\end{eqnarray}
where we did not consider the small $t_b$ dependent term in equation (\ref{total-EE-EGB}). Location of the island surface can be found by extremizing the above equation with respect to $a$, i.e.,
\begin{eqnarray}
\label{der-sgen-a-EGB}
\frac{\partial S_{\rm gen}(a) }{\partial a}\sim  \frac{4 \pi  r_+}{G_N}+\frac{c \left(\frac{r_+-r_-}{b-r_-}\right){}^{-\frac{r_-^2}{r_+^2}} e^{\kappa _+ \left(r_+-b\right)}}{3 \left(r_+-b\right) \sqrt{\frac{a-r_+}{b-r_+}}} =0,
\end{eqnarray}
equation (\ref{der-sgen-a-EGB}) has the following solution,
\begin{eqnarray}
\label{a-GB}
& & 
a \approx r_+ +\frac{c^2 \left(\frac{r_+-r_-}{b-r_-}\right){}^{-\frac{2 r_-^2}{r_+^2}} G_N^2 e^{2 \kappa _+ \left(r_+-b\right)}}{144\pi ^2 r_+^2 \left(b-r_+\right)}.
\end{eqnarray}

Now substituting value of $a$ from equation (\ref{a-GB}) in equation (\ref{total-EE-EGB}), total entanglement entropy of the charged Einstein-Gauss-Bonnet black hole at late times (\ref{total-EE-EGB}) simplifies to the following form \footnote{As explained in footnote \ref{Smatter-footnote} that matter contribution, $S_{\rm mater}({\cal R} \cup {\cal I})$, has different exponential factor in our case.}
:
\begin{eqnarray}
\label{total-EE-EGB-simp}
& &
S_{\rm total}^{\rm EGB}\sim \frac{2 \pi  \left(4 \alpha +r_+^2\right)}{G_N}+\frac{c}{3}  \log \left(\frac{r_- e^{- \kappa _+ \left(b+r_+\right)} \left(\frac{r_-^2}{\left(r_+-r_-\right) \left(b-r_-\right)}\right){}^{\frac{1}{2} \left(\frac{\kappa _+}{\kappa _-}-1\right)}}{b
   \kappa _+^2}\right),\nonumber\\
   & & S_{\rm total}^{\rm EGB}  = 2 S_{BH}^{(\alpha)} +\frac{c}{3}  \log \left(\frac{r_- e^{-\kappa _+ \left(b+r_+\right)} \left(\frac{r_-^2}{\left(r_+-r_-\right) \left(b-r_-\right)}\right){}^{\frac{1}{2} \left(\frac{\kappa _+}{\kappa _-}-1\right)}}{b
   \kappa _+^2}\right).
\end{eqnarray}
Since for the charged black hole in Einstein-Gauss-Bonnet gravity
total entanglement entropy at late times (\ref{total-EE-EGB-simp}) is same as total entanglement entropy for the charged black hole in higher derivative gravity with ${\cal O}(R^2)$ terms given in equation (\ref{S-total-HD}). Therefore Page curves of the charged black hole in Einstein-Gauss-Bonnet gravity will be same as the Page curves of charged black hole in higher derivative gravity with ${\cal O}(R^2)$ terms (figure {\ref{SEE-versus-alpha})}. \par 
{\bf Page Time:}
Since at $t=t_{\rm Page}^{\rm EGB}$, $S_{\rm EE}^{\rm WI}=S_{\rm total}^{\rm EGB}$, therefore on equating (\ref{EE-ET-ET-1}) and (\ref{total-EE-EGB}), we obtain the Page time of the charged Einstein-Gauss-Bonnet black hole as given below:
\begin{eqnarray}
\label{Page-time-EGB}
t_{\rm Page}^{\rm EGB}=\frac{6 \pi  \left(4 \alpha +r_+^2\right)}{c \kappa _+ G_N} = \frac{3 S_{\rm BH}^{(0)}}{2 \pi c T_{\rm RN}}+\frac{12 \alpha}{c \ G_N T_{\rm RN}}=t_{\rm Page}^{(0)}+\frac{12 \alpha}{c \ G_N T_{\rm RN}}.
\end{eqnarray}
From equations (\ref{total-EE-EGB}) and (\ref{Page-time-EGB}) we can see that in the $\alpha \rightarrow 0$ limit we obtain the total entanglement entropy at late times and Page time of charged black hole without higher derivative term as reviewed in section \ref{review-RNBH-Page-curve}. Physical significance of the Gauss-Bonnet term is that when Gauss-Bonnet coupling is increasing then we are getting the Page curves at later time and when Gauss-Bonnet coupling is decreasing then we are getting the Page curves at earlier time with reference to Page curve of charged black hole without higher derivative terms, i.e., Page curve of the Reissner-Nordstr{\"o}m black hole. \par

\par
{\bf Scrambling time:} In this case scrambling time is unaffected by the higher derivative terms and is same as scrambling time of the Reissner-Nordstr{\"o}m black hole \cite{Island-RNBH},
\begin{eqnarray} 
\label{tscr-GB-term}
t_{scr}^{\rm EGB}\sim \frac{2 r_+^2}{(r_+ - r_-)} \log\left(\frac{\pi r_+^2}{G_N}\right) + small = \frac{1}{2 \pi T_{\rm RN}}  \log\left(S_{\rm BH}^{(0)}\right) + small=t_{\rm scr}^{(0)}+small.
\end{eqnarray}

\section{Conclusion and Discussion}
\label{Summary}
In this paper we have obtained the Page curves of an eternal Reissner-Nordstr{\"o}m black hole in the presence of higher derivative terms where entanglement entropy had been calculated using the formula given in \cite{Dong}. \par
Island rule was originated by coupling the evaporating JT(Jackiw Teitelboim) black hole plus conformal matter to two dimensional CFT bath \cite{AMMZ}. Idea was to consider a black hole in JT gravity plus conformal matter (JT gravity + 2D CFT) to a CFT bath. There are three equivalent descriptions of this setup.
\begin{itemize}
\item {\bf 2D-Gravity:} Black hole in JT gravity plus matter theory is coupled to 2D CFT bath where we are collecting the Hawking radiation.
\item {\bf 3D-Gravity:} The CFT bath has its own holographic dual which is $AdS_3$.
\item {\bf QM:} Boundary of JT gravity plus matter theory is one dimensional QM system. Third description is 2D CFT bath with some boundary QM degrees of freedom. 
\end{itemize} 

Island rule for evaporating black holes in the above setup arises in a beautiful way that extra dimension(radial coordinate) in $AdS_3$ background connects the CFT bath to a region in the interior of the black hole which is known as, ``Island", which appear at late times. Initially there is no island and one finds that entanglement entropy of Hawking radiation is increasing linearly with time and after the Page time island comes into the picture (which gives finite contribution to the entanglement entropy) and we obtain the Page curve. Above discussion depends strongly on holographic duality.\par

Island formula can be derived from gravitational path integral using replica trick for special JT black holes as done in \cite{rw-1,rw-2}. It was discussed by authors that one obtains the Page curve of eternal black holes in the following way. There are two saddles: disconnected and connected. Below the Page time disconnected saddle dominates and after the Page time connected saddle dominates which is replica wormhole. Disconnected saddle is responsible for the linear time growth of entanglement entropy of the Hawking radiation and connected saddle gives the finite contribution. Combination of contributions from the disconnected and connected saddles to the entanglement entropy of Hawking radiation reproduces the Page curve. The argument of \cite{rw-1} is also applicable to $n$ boundary wormholes which are known as replica wormholes.\par

In this paper we are focusing on non-holographic models therefore we should be careful when we will be discussing the higher dimensional asymptotically flat black holes. Since we are assuming that observer is very far away from the black hole therefore we can use s-wave approximation to calculate the entanglement entropy of Hawking radiation using formula for 2D CFT. Most of the papers existing in the literature discuss the application of island proposal to higher dimensional asymptotically flat black holes for the eternal black holes only. There is a paper \cite{NBH-HD} in which authors have discussed the Page curve in higher dimensional evaporating black hole in the presence of higher derivative terms in the gravitational action.\par

Effect of higher derivative terms on the Page curve had been explored in \cite{NBH-HD} for the neutral black hole. Authors considered the eternal black hole as well as evaporating black hole in their study. They focus only on the general ${\cal O}(R^
2)$ terms in the gravitational action as higher derivative terms. We consider the non-extremal Reissner-Nordstr\"om black hole and we are focusing only on the eternal black hole. It is nice to study the effect of the Gauss-Bonnet coupling on the Page curves of Reissner-Nordstr\"om black hole. Presence of higher derivative terms in the gravitational action affect the Page time, scrambling time and Page curves of eternal black holes.
Following are the main results obtained in this paper.
\begin{enumerate}
\item In the first case we considered the general ${\cal O}(R^2)$ terms in the gravitational action as discussed in \cite{NBH-HD} and calculated the Page curves. We find that, since initially there is no island surface, the entanglement entropy of the Hawking radiation increases linearly with time forever. Therefore we have information paradox for the charged black hole in the presence of ${\cal O}(R^2)$ terms. At late times island emerges  
and entanglement entropy of the Hawking radiation reaches a constant value which is twice of the Bekenstein-Hawking entropy of the black hole and we obtain the Page curves for fix values of the Gauss-Bonnet coupling ($\alpha$). In this case we find that {\it as Gauss-Bonnet coupling ($\alpha$) increases Page curves shift towards later times and when Gauss-Bonnet coupling ($\alpha$) decreases Page curves shift towards earlier times.}

\item In the second case we considered only Gauss-Bonnet term as higher derivative term which is relevant to studying the charged black hole in Einstein-Gauss-Bonnet gravity \cite{Charged-GB-BH}. Similar to the first case we calculated the Page curves in this setup, i.e., of charged Einstein-Gauss-Bonnet black hole \cite{Island-RNBH}. Similar to the first case we have linear time growth of the entanglement entropy of the Hawking radiation at earlier times and island emerges at late times which saturates the linear time growth of the Hawking radiation and entanglement entropy of Hawking radiation becomes equal to twice of the Bekenstein-Hawking entropy of the black hole. Therefore we obtain the Page curves of the charged Einstein-Gauss-Bonnet black hole in four dimensions. Interestingly we find a similar behaviour of the Page curves with Gauss-Bonnet coupling as discussed earlier in the first case.

\item In the first case scrambling time of the charged black hole is affected by the higher derivative terms. It will increase if the correction term is positive and decrease if the correction term is negative. Further if we take the coupling appearing in scrambling time to zero then we recover the scrambling time of the Reissner-Nordstr{\"o}m black hole. In the second case, i.e., when we consider only Gauss-Bonnet term as higher derivative term, scrambling time is unaffected by the higher derivative term and is same as scrambling time of the Reissner-Nordstr{\"o}m black hole.

\item In the limit of vanishing Gauss-Bonnet coupling, we obtain the Page curve of the Reissner-Nordsrt{\"o}m black hole as obtained in \cite{Island-RNBH}. 
\end{enumerate}

\section*{Acknowledgements}
The author is supported by a Senior Research Fellowship (SRF) from the Council of Scientific and Industrial Research, Govt. of India. The author was benefited from{\it "Kavli Asian Winter School (KAWS) on Strings, Particles and Cosmology(Online)" (code: ICTS/kaws2022/1)}. The author would like to thank Aalok Misra for helpful discussions. The author would also like to thank Xuanhua Wang, Xian-Hui Ge and Mohsen Alishahiha for various correspondences and clarifications.

\end{document}